\documentstyle[12pt]{article}

\catcode`\@=11
\def\marginnote#1{}
\newcount\hour
\newcount\minute
\newtoks\amorpm
\hour=\time\divide\hour by60
\minute=\time{\multiply\hour by60 \global\advance\minute by-\hour}
\edef\standardtime{{\ifnum\hour<12 \global\amorpm={am}%
        \else\global\amorpm={pm}\advance\hour by-12 \fi
        \ifnum\hour=0 \hour=12 \fi
        \number\hour:\ifnum\minute<10 0\fi\number\minute\the\amorpm}}
\edef\militarytime{\number\hour:\ifnum\minute<10 0\fi\number\minute}
\def\draftlabel#1{{\@bsphack\if@filesw {\let\thepage\relax
   \xdef\@gtempa{\write\@auxout{\string
      \newlabel{#1}{{\@currentlabel}{\thepage}}}}}\@gtempa
   \if@nobreak \ifvmode\nobreak\fi\fi\fi\@esphack}
        \gdef\@eqnlabel{#1}}
\def\@eqnlabel{}
\def\@vacuum{}
\def\draftmarginnote#1{\marginpar{\raggedright\scriptsize\tt#1}}
\def\draft{\oddsidemargin -.5truein
        \def\@oddfoot{\sl preliminary draft \hfil
        \rm\thepage\hfil\sl\today\quad\militarytime}
        \let\@evenfoot\@oddfoot \overfullrule 3pt
        \let\label=\draftlabel
        \let\marginnote=\draftmarginnote
   \def\@eqnnum{(\theequation)\rlap{\kern\marginparsep\tt\@eqnlabel}%
\global\let\@eqnlabel\@vacuum}  }

\def\preprint{\twocolumn\sloppy\flushbottom\parindent 1em
        \leftmargini 2em\leftmarginv .5em\leftmarginvi .5em
        \oddsidemargin -.5in    \evensidemargin -.5in
        \columnsep 15mm \footheight 0pt
        \textwidth 250mmin      \topmargin  -.4in
        \headheight 12pt \topskip .4in
        \textheight 175mm
        \footskip 0pt
        \def\@oddhead{\thepage\hfil\addtocounter{page}{1}\thepage}
        \let\@evenhead\@oddhead \def\@oddfoot{} \def\@evenfoot{} }

\def\titlepage{\@restonecolfalse
\if@twocolumn\@restonecoltrue\onecolumn
     \else \newpage \fi \thispagestyle{empty}\c@page\z@
        \def\thefootnote{\fnsymbol{footnote}} }

\def\endtitlepage{\if@restonecol\twocolumn \else  \fi
        \def\thefootnote{\arabic{footnote}}
        \setcounter{footnote}{0}}  

\catcode`@=12
\relax
\def\bea{\begin{array}}
\def\bem{\begin{displaymath}}
\def\beq{\begin{equation}}
\def\eea{\end{array}}
\def\eem{\end{displaymath}}
\def\eeq{\end{equation}}

\def\NP#1#2#3{Nucl. Phys. \underline{#1} (19#2) #3}

\def\PL#1#2#3{Phys. Lett. \underline{#1} (19#2) #3}
\def\PR#1#2#3{Phys. Rev. \underline{#1} (19#2) #3}

\def\Re{\mathop{\rm Re}}

\relax
\newcommand{\be}{\begin{equation}}
\newcommand{\en}{\end{equation}}
\newcommand{\ba}{\begin{eqnarray}}
\newcommand{\ea}{\end{eqnarray}}
\newcommand{\ee}{\end{equation}}
\def\alp{\alpha^\prime}

\def\crbig{\\\noalign{\vspace {3mm}}}

\relax

\begin{document}
\topmargin-2.4cm
\renewcommand{\theequation}{\thesection.\arabic{equation}}
%
%
%
%
\begin{titlepage}
\begin{flushright}
CERN--TH/99-31\\
LPTENS-99/03\\
hep--th/9902072\\
February 1999
\end{flushright}
\vspace{.1cm}
\begin{center}{\large  \bf Universal  Thermal Instabilities
and the High-\\
\vskip .1cm
Temperature Phase of the  ${\bf N=4}$
Superstrings$^{\star}$}
\vskip .5cm
{\bf Costas Kounnas}
\vskip .1cm
{\normalsize\sl LPTH, Ecole Normale
Sup\'erieure, 24 rue Lhomond,\\
F-75231 Paris Cedex 05, France,\\
\vskip .1cm
and}
\vskip .1cm
{\normalsize\sl  Theory Division, CERN, 1211 Geneva 23,
Switzerland}
\end{center}

\vskip .2cm
\begin{center}
{\bf Abstract}
\end{center}
\begin{quote}
Using the properties of gauged $N=4$ supergravity, we show that it is
possible to derive a universal thermal effective potential that
describes all possible high-temperature instabilities of the known
$N=4$ superstrings. These instabilities are due to non-perturbative
dyonic modes, which become tachyonic in a region of the
thermal moduli space ${\rm \cal M}=\{s,t,u\}$; ${\rm \cal M}$ is
common to all non-perturbative dual-equivalent $N=4$ superstrings in
five dimensions. We analyse the non-perturbative thermal potential
and show the existence of a phase transition at high temperatures
corresponding to a condensation of 5-branes. This phase is
described in detail, using an effective non-critical string theory.
\end{quote}
\vspace*{.1cm}
\begin{center}
{\it Talks given at the 6th Hellenic School and Workshops}\\
{\it Corfou, 6-26 Septemper 1998} \\
{\it ``Quantum aspects of gauge theories, supersymmetry and unification"}\\
{\it TMR mid-term meeting, 19-26 Septemper 1998}

\end{center}

\vspace{.1cm}
\begin{flushleft}
\rule{8.1cm}{.2mm}\\
$^{\star}$
{\small Research supported in part by
the EEC under the TMR contract
ERBFMRX-CT96-0045.\\
e-mail: kounnas@nxth04.ch}
\end{flushleft}
\end{titlepage}
\setcounter{page}{0}
\setlength{\baselineskip}{.7cm}
\newpage
%
%
\section{Introduction}\label{secintro}
\setcounter{equation}{0}

At finite temperatures the partition function $Z(\beta)$ and the mean
energy $U(\beta)$ develop power pole singularities in $\beta\equiv
T^{-1}$ if the density of states of a system grows exponentially with
the energy:
$$
\rho(E)\sim E^{-k}~e^{bE}\, ,
$$
$$
Z(\beta) =\displaystyle{\int dE~\rho(E)~ e^{-\beta E}
\sim {1\over{(\beta-b)}^{(k-1)}}\, },
$$
\beq
U(\beta) = \displaystyle{ -{\partial\over\partial\beta}\ln Z
\sim (k-1){1\over\beta-b}+...}
\eeq
At the critical temperature, $ T_H=b^{-1}$, various thermodynamical
quantities diverge \cite{H}.
An alternative interpretation of the pole singularity in $~U(\beta)$
follows from the identification of the temperature with the inverse
radius of a compactified Euclidean time on $S^1$: $2\pi T =1/R $. In
this representation, the partition function is given by the
(super-)trace over the thermal spectrum of the theory in $(D-1)$
dimensions:
\beq
\ln Z(\beta)={\rm Str}\ln{\cal M}(\beta).
\eeq
The pole singularity is then a manifestation of a thermal state
becomes massless at the Hagedorn temperature $T_H$. Thus, the
knowledge of the thermal spectrum ${\cal M}(\beta)$ as a function of
the $S^1$ radius $R=\beta/2\pi$, determines $T_H$ \cite{S}.

Perturbative string theories provide examples of an exponentially
growing density of states, with $k=D$, the dimension of space-time,
and $b^{-1}\sim $ ${\cal O}(\alpha^\prime)^{-1/2}$
\cite{A}--\cite{AxK}.
In superstrings the states that become tachyonic at $T_H$ have
necessarily a non-zero winding number $n$ \cite{ AW,KR,AK}.

{From the perturbative study of the $N=4$ strings we can see that
the
states that become tachyonic above $T_H$ correspond to the $N=4$ BPS
states that preserve half of the supersymmetries ($N=2$) \cite{ADK}.
However, in the $N=4$ theories the the masses of the non-perturbative
BPS states are known as well thanks to the $N=4$ supersymmetry
algebra with
central extension \cite{N=4BPS}--\cite{KK}. We can therefore identify
all perturbative and non-perturbative BPS states that are able to
induce high-$T$ instabilities using the string duality properties
among the heterotic--type IIA--type IIB--type I strings with $N=4$
supersymmetry \cite{HT}--\cite{sixdimdual}.

In this talk I will summarize the results of Ref. \cite{ADK}
concerning
non-perturbative $N=4$ theories at finite temperatures.

\vskip .4cm
\section{ Thermally  modified  $N=4$ BPS masses}
\vskip .1cm

In order to obtain the thermal partition function one modifies the
boundary conditions around the $S^1$-Euclidean time by a
spin-statistical phase. In perturbative string theories the
consequence of this phase is to shift the Kaluza--Klein momenta of
the
$S^1$
\beq
P_{L,R}=\left(~{m + Q -{n \delta \over 2}\over R}\pm
{n R \over \alp }\right)^2,
\eeq
and reverse the GSO projection in the $n$-odd winding sector;
$Q=Q_L+Q_R$ is the helicity operator while $\delta=1$ in the
heterotic
string and $\delta=0$ in type II strings \cite{AW,KR,AK,ADK}.

The left- (and right- ) supersymmetric GSO projection(s) implies that
in the even winding sector all states have $M^2\ge 0$. However, some
of the states with odd winding number can become tachyonic by a
reversion of the GSO-projection \cite{AW, KR, AK, ADK}. The only
states that can become tachyonic have $n=\pm 1$ and left-helicity
$Q_L=\pm 1$ (right-helicity $=-Q_R$ for type II). They are scalars in
$(D-1)$-dimensions (the longitudinal components of the
$D$-dimensional gravitons). The Hagedorn temperature corresponds to
the critical value of the $S^{1}$ radius at which the first tachyonic
state appears when $2\pi R=T^{-1}$ decreases.

The appearance of tachyons can never arise in any perturbative
supersymmetric field theory, since it behaves like the zero-winding
sector of strings. In non-perturbative supersymmetric field theories,
however, such an instability can arise due to thermal dyonic modes,
which behave like the odd winding string states. Indeed, before the
temperature modification, the heterotic--type II duality in five
dimensions exchanges the winding number $n$ with the magnetic charge
$\ell$:
\beq
\label{shift}
{\cal M}^2 =
\left( {m\over R} + {nR\over\alp_H} + {\ell
R\over\lambda_H^2
\alp_H} \right)^2,
\eeq
where $m$ and $n$ are the $S^1$ momentum and winding numbers, and
$\ell$ is the
non-perturbative wrapping number for the heterotic 5-brane around
$T^4\times S^1$; $\lambda_H$ is the string coupling in $D=6$; the
tension of the 5-brane $T_5$=${1/\lambda_H^{2}}$ in $\alp_{H}$
units. Using the $\cal S$-duality relations:
\beq \lambda_H =
{1\over\lambda_{IIA}},~~~~ ~~~~~~ \lambda_H^2 \alpha^\prime_H =
\alpha^\prime_{II} ,
\eeq
we can express the above  mass formula
in terms of type IIA parameters:
\beq
{\cal M}^2 = \left( {m\over R} +
{nR\over\alpha^\prime_{II}\lambda_{IIA}^2}
+  {\ell R\over\alp_{II}} \right)^2=\left( {m\over R} +
{nR\over\alpha^\prime_{H}}
+  {\ell R\over\alp_{II}} \right)^2.
\eeq
The momentum and winding numbers are now {$m$} and $\ell$; $n$ is
the wrapping number for the type IIA NS 5-brane around $K_3\times
S^1$. From the six-dimensional viewpoint, $m/R$ is the Kaluza--Klein
momentum, while the last two terms correspond to BPS strings with
tension
\beq
T_{p,q}={p\over \alp_H}+{q\over\alp_{II}}\, ,
\eeq
where $p,q$ are relatively prime integers, $(n,l)=k~(p,q)$.
The common divisor $k$
defines the wrapping of the $T_{p,q}$ string around $S^1$; $q$ is the
charge of the fundamental string and $p$ the magnetic charge of the
solitonic string obtained by wrapping the NS 5-brane around $K_3$.
The $T_{p,q}$-strings cannot become tensionless since thet
never are associated to vanishing cycles of the internal manifold.

The five-dimensional thermal mass formula is obtained by the
non-perturbative
generalization of the temperature deformation using the
$(p,q)$-string picture of the non-perturbative BPS spectrum by
replacing $m \rightarrow m+Q'+{n\over2}$ and reversing the
GSO-projection in the $k$-odd sector of the $(p,q)$-strings ($Q'$ is
the
helicity operator of the 5D-thermal theory):
\beq
\label{mass3}
{\cal M}^2_T = \left( {m+Q'+{kp\over2}\over R} +
k~T_{p,q}~R \right)^2 - 2 ~T_{p,q}~ \delta_{k,\pm1} ~\delta_{Q',0}\,.
\eeq
This thermal formula reproduces the perturbative result for both
heterotic and type IIA theories. In the heterotic perturbative limit
$\lambda_H\rightarrow 0$, only the $\ell=0=q$ states survive, while
in the type IIA perturbative limit $\lambda_{II}\to 0$, only the
$n=0=p$ states survive. Note that in the general case of a $T_{p,q}$
string with the temperature deformation, the condition $mk\ge 0$
becomes $mk\ge-1$, because of the inversion of the GSO-projection.

{}From Eq. (\ref{mass3}), it follows  that if the heterotic coupling
$\lambda_H$ is smaller than the critical value
\beq
\label{crit}
\lambda_H ~<~\lambda_H^c = {\sqrt2+1\over2},
\eeq
the first tachyon appears at $R=(\sqrt2+1)\sqrt{\alpha^\prime_H/2}$,
which corresponds to the heterotic Hagedorn temperature. On the other
hand, if the heterotic theory is strongly coupled,
$\lambda_H>\lambda_H^c$, the first tachyon appears at
$R=\sqrt{2\alp_H}\,\lambda_H = 2\sqrt{\alp_{II}/2}$; this corresponds
to the type IIA Hagedorn temperature. Besides the above two would-be
tachyons, the mass formula (\ref{mass3}) leads in general to two
series
of potentially tachyonic states with $m=-1$. However the critical
temperature $(2\pi R)^{-1}$ for each of the states in both series is
always higher than the lowest Hagedorn heterotic temperature while,
as discussed above, the type IIA Hagedorn temperature first appears
when the heterotic coupling exceeds the critical value $\lambda_H^c$
(\ref{crit}).

In order to include the type IIB dual $N=4$ string, we need to
discuss five-dimensional theories at finite temperature, taking into
account the compactification radius $R_6$ from six to five
dimensions. Type IIA and IIB strings are then related by the
inversion of
$R_6$. The extension to four dimensions of the mass formula
(\ref{mass3}) is straightforward. It depends on three parameters, the
string coupling $g_H$, the temperature radii $R$ and $R_6$. It is
convenient to introduce the three combinations
\beq
\label{stuare}
t = {RR_6\over\alpha^\prime_H},
\qquad u = {R\over R_6}, \qquad  s= g_H^{-2} = {t\over\lambda_H^2},
\eeq
in terms of which the thermally shifted BPS mass formula
reads \cite{ADK}:
\beq
\label{mass5}
{\cal M}^2_T =
\left({m+Q'+{kp\over 2}\over R}+
k~T_{p,q,r}~R\right)^2-2 ~T_{p,q,r}~\delta_{|k|,1}
{}~\delta_{Q',0} ~,
\eeq
where
$T_{p,q,r}$ is then an effective string tension
\beq
\label{Tpqr}
T_{p,q,r}={p\over\alpha_H^\prime}
+{q\over\lambda_H^2\alpha_H^\prime}
+{r R_6^2\over\lambda_H^2(\alpha_H^\prime)^2}={p\over\alpha_H^\prime}
+{q\over\alpha_{IIA}^\prime}
+{r\over\alpha_{IIB}^\prime}~ ;
\eeq
here, the various $\alpha^\prime$ and the radius $R$ are expresed in
terms of $s,t,u$ and in termes of the four-dimensional Planck scale
$\kappa=\sqrt{8\pi}M_P^{-1}$:
\beq
\label{alphaare}
\alpha_H^\prime = 2\kappa^2 s, \qquad
\alpha_{IIA}^\prime = 2\kappa^2 t, \qquad
\alpha_{IIB}^\prime = 2\kappa^2 u, \qquad
R^2 = \alpha^\prime_H tu  = 2\kappa^2 stu
\eeq
The temperature radius $R$ is by construction identical in all three
string theories. Note that $l=kq$ corresponds to the wrapping number
of the heterotic 5-brane around $T^4\times S^1_R$ as in five
dimensions, while $l^{\prime}=kr$ corresponds to the same wrapping
number after performing a $\rm \cal T$-duality along the $S^1_{R_6}$
direction, which is orthogonal to the 5-brane. All winding numbers,
$n,l,l'$, correspond to magnetic charged states from the
field theory point of view. Their masses are proportional to the
temperature-radius $R$ and are not thermally shifted; $p,q,r$ are all
non-negative relatively prime integers. This follows from the
BPS conditions and the $s\leftrightarrow t\leftrightarrow u$ duality
symmetry in the undeformed supersymmetric theory. Futhermore, $mk\ge
-1$ because of the inversion of the GSO projection in the
temperature-deformed theory. Using these constraints, one can show
that there are, in general, two potential tachyonic series with
$m=-1$
and $p=1,2$ \cite{ADK} generalizing the five-dimensional result.
One of the perturbative heterotic, type IIA, or type IIB potential
tachyons corresponds to a critical temperature that is always lower
than those of the two series with $p=1,2$.

The above discussion shows that the temperature modification of the
mass formula inferred from perturbative strings and applied to the
non-perturbative BPS mass formula produces the appropriate
instabilities in terms of the Hagedorn temperatures. In Ref.
\cite{ADK} we show that it is possible to go beyond the simple
enumeration of Hagedorn temperatures and construct the full
temperature-dependent effective potential associated with the
would-be tachyonic states. This will allow a study of the nature of
the non-perturbative instabilities and the dynamics of the various
thermal phases.

\section{ Thermal effective potential}
\vskip .1cm

Our procedure to construct the thermal effective theory is as
follows:
we consider five-dimensional $N=4$ theories at finite temperature.
They can then effectively be described by four-dimensional theories,
in which supersymmetry is spontaneously broken by thermal effects.
Since we want to limit ourselves to the description of instabilities,
it is sufficient to retain, in the full $N=4$ spectrum, only  the
potentially massless and tachyonic states. This restriction will lead
us to consider only spin 0 and 1/2 states, the graviton and one of
the gravitinos. This sub-spectrum is described by an $N=1$
supergravity with six chiral multiplets: the three moduli $S,~T,~U$,
and the three would-be tachyonic states $Z_A,~A=1,2,3.$ Using the
properties of the $N=4$ (gauged) supergravities in four dimensions
\cite{DF}--\cite{dR},\cite{AK}, it is possible to derive the
temperature motification associated to the Scherk-Schwarz temperature
gauging \cite{susybreak, PZ,AK, KR}. This is done in Ref. \cite{ADK},
where the K\"ahler manifold $K$ and the superpotential $W$ of the
effective $N=1$ supergravity \cite{CFGVP} are derived:
$$
\label{Kis}
K = -\log(S+S^*)-\log(T+T^*)-\log(U+U^*)
-2\log (1 -2\vert Z_A \vert^2 + \vert Z_A^2\vert^2) ~,
$$
\beq
\label{Wis}
W = 2\sqrt2 \biggl[ {1\over 2}(1-Z^2_A)^2
+ (TU-1)Z_1^2  + SU Z_2^2 +
STZ_3^+Z_3^2\biggr] ~.
\eeq
The resulting scalar potential has a complicated expression.
Important
simplifications occur, however, when we restrict ourselves to the
would-be tachyonic states, $z_A=\Re Z_A$ and in $s=\Re S$, $t=\Re T$,
$u=\Re U$. Following the analysis of Ref. \cite{ADK} only these
directions are relevant to the vacuum structure of the potential and
to possible phase transitions. The resulting scalar potential $V$ is:
\beq
\label{pot4}
\begin{array}{rcl}
V &=& V_1 + V_2 + V_3, \crbig
\kappa^4 V_1 &=& \displaystyle{4\over s}\left[
(\xi_1+\xi_1^{-1})H_1^4
+{1\over4}(\xi_1-6+\xi_1^{-1})H_1^2 \right],
\crbig
\kappa^4 V_2 &=& \displaystyle{4\over t}
\left[ \xi_2H_2^4 +{1\over4}(\xi_2-4)H_2^2\right],
\crbig
\kappa^4 V_3 &=& \displaystyle{4\over u}
\left[ \xi_3H_3^4 +{1\over4}(\xi_3-4)H_3^2\right],
\end{array}
\eeq
where the moduli fields $\xi_i$ are given in terms of $R^2$ and the
various $\alpha^\prime$:
\beq
\label{xidef}
\xi_1 = tu={2R^2\over\alpha^\prime_H},
\qquad \xi_2 = su={2R^2\over\alpha^\prime_{IIA}},
\qquad \xi_3 = st={2R^2\over\alpha^\prime_{IIB}}.
\eeq
$V$ is a simple fourth-order polynomial, when expressed in terms of
new
field variables $H_A,~A=1,2,3$,
\beq
H_A={z_A\over (1-x^2)}~,~~~~x^2=\left( 1-\sum_A Z_A^2\right)~,
\eeq
that take values on the entire real axis.
At $H_i=0$, the K\"ahler metric is $4\delta^A_B$, the scalar
potential is normalized according to $V = {4\kappa^2}\sum_A m_A^2
H_A^2 + \ldots$. The masses $m_A^2$ correspond to the
mass formula for the heterotic, IIA and IIB tachyons.

Having the full effective thermal potential of the theory we able to
study the phase structure of the thermal effective theory. There are
four different phases corresponding to specific regions of the $s$,
$t$ and $u$ moduli space. Their boundaries are defined by critical
values of the moduli $s$, $t$, and $u$ (or of $\xi_i$, $i=1,2,3$), or
equivalently by critical values of the temperature, the
(four-dimensional) string coupling and the compactification radius
$R_6$. These four phases are \cite{ADK}:
\begin{enumerate}
\item The {\sl low-temperature} phase:

$T<(\sqrt2-1)^{1/2}/(4\pi\kappa)\,$.

\item The {\sl high-temperature heterotic} phase:

$T>(\sqrt2-1)^{1/2}/(4\pi\kappa)\,\,$ and
$\,\,g_H^2<(2+\sqrt2)/4\,$.

\item The {\sl high-temperature type IIA} phase:

$T>(\sqrt2-1)^{1/2}/(4\pi\kappa)$\,\,, $\,\,g_H^2>(2+\sqrt2)/4\,\,$
and $\,\,R_6>\sqrt{\alpha^\prime_H}\,$.

\item The {\sl high-temperature type IIB} phase:

$T>(\sqrt2-1)^{1/2}/(4\pi\kappa)\,\,$, $\,\,g_H^2>(2+\sqrt2)/4\,\,$
and $\,\,R_6<\sqrt{\alpha^\prime_H}\,$.
\end{enumerate}

$\bullet$ The {\sl low-temperature phase}, which is common to all
three
strings, is characterized by
\beq
\label{ltp1}
H_1=H_2=H_3=0, \qquad V_1=V_2=V_3=0.
\eeq
The potential vanishes for all values of the moduli $s$, $t$ and $u$,
which are then only restricted by the stability of the phase, namely
the absence of tachyons in the mass spectrum of the scalars $H_A$.
Since the (four-dimensional) string couplings are
$$
s = \sqrt2 g_H^{-2}, \qquad t = \sqrt 2 g_A^{-2}, \qquad
u = \sqrt2 g_B^{-2},
$$
this phase exists in the perturbative regime of all three strings.
The relevant light thermal states are just the massless modes of the
five-dimensional $N=4$ supergravity, with thermal mass scaling like
$1/R \sim T$.
\vskip .1cm
$\bullet$ The {\sl high-temperature heterotic phase}
is defined by
\beq
\label{hhp1}
\xi_H > \xi_1 > {1\over\xi_H},
\qquad\qquad \xi_2>4, \qquad\qquad \xi_3>4,
\eeq
with $\xi_H =(\sqrt2+1)^2$. The inequalities on $\xi_2$ and $\xi_3$
eliminate type II instabilities. In this region of the moduli, the
potential becomes, after minimization with respect to $H_1$, $H_2$
and $H_3$:
$$
\kappa^4 V = -{1\over s}{(\xi_1+\xi_1^{-1}-6)^2\over
16(\xi_1+\xi_1^{-1})}.
$$
It has a stable minimum for fixed $s$ (for fixed $\alpha^\prime_H$)
at the minimum of the self-dual quantity $\xi_1+\xi_1^{-1}$:
\beq
\label{highhetmin}
\xi_1=1 , \qquad H_1={1\over2} , \qquad H_2=H_3=0, \qquad
\kappa^4 V= -{1\over 2s}.
\eeq
The transition from the low-temperature vacuum is due to a
condensation of the heterotic thermal winding mode $H_1$ or,
equivalently, to a condensation of type IIA NS 5-brane in the type
IIA picture. At the level of the potential only, this phase exhibits
a runaway behaviour in $s$. We will show in the next section that a
stable solution to the effective action exists with non-trivial
metric and/or dilaton.

\vskip .1cm
$\bullet$ The {\sl high-temperature type IIA and IIB phases}
are defined by the inequalities
\beq
\label{IIphases1}
\xi_2<4 \qquad {\rm and/or}\qquad \xi_3<4.
\eeq
In this region of the parameter space, either
$H_2$ or $H_3$ become tachyonic and acquire a vacuum value:
\beq
\label{IIAphase}
H_2^2 = {4-\xi_2\over 8\xi_2}, \qquad\qquad
\kappa^4 V_2=-{1\over t}{(4-\xi_2)^2\over16\xi_2},
\eeq
and/or
\beq
\label{IIBphase}
H_3^2 = {4-\xi_3\over 8\xi_3}, \qquad\qquad
\kappa^4 V_3=-{1\over u}{(4-\xi_3)^2\over16\xi_3}.
\eeq
In contrast with the high-temperature heterotic phase, the potential
does not possess stationary values of $\xi_2$ and/or $\xi_3$, besides
the critical points $\xi_{2,3}=4$. Suppose for instance that
$\xi_2<4$
and $\xi_3>4$. The resulting potential is then $V_2$ only and $\xi_2$
slides to zero. In this limit, $ V= -\,{1/( stu\kappa^4)}, $
and the dynamics of $\phi \equiv -\log(stu)$ is described by the
effective Lagrangian
$$
{\cal L}^{II}_{\rm eff} = -{e\over2\kappa^2}\left[R
+{1\over6}(\partial_\mu\phi)^2 - {2\over\kappa^2}e^{\phi}\right].
$$
The other scalar components $\log(t/u)$ and $\log(s/u)$ have only
derivative couplings since the potential only depends on $\phi$. They
can be taken as constant and arbitrary. The dynamics only restricts
the
temperature radius $\kappa^{-2}R^2=e^{-\phi}$, $R_6$ and the string
coupling is not constrained. The ground state of this phase
corresponds to a non-trivial gravitational and dilaton background
satisfying the  Einstein and $\phi$ equations of ${\cal
L}^{II}_{\rm eff}$. This background solution defines the
high-temperature type II vacuum. We will not study this
solution further. Instead, we will examine in more detail the
high-$T$
heterotic phase.

\section{The high-$T$ heterotic phase transition}
\vskip .1cm

Using the information contained in the effective theory, which is
characterized by Eqs. (\ref{highhetmin}), the equations of motion of
all
scalar fields are satisfied, with the exception of the dilaton $s=\Re
S$.
The resulting bosonic effective Lagrangian that describes the
dynamics
of $s$ and $g_{\mu\nu}$ is:
\beq
\label{hethigh1}
{\cal L}^H_{\rm bos} = -{1\over2\kappa^2}eR - {e\over4\kappa^2}
(\partial_\mu\ln s)^2 + {e\over2\kappa^4s}.
\eeq
For all (fixed) values of $s$, the cosmological constant is negative,
since $V=-(2\kappa^4s)^{-1}$ and the apparent geometry is
anti-de Sitter. But the effective theory (\ref{hethigh1}) does not
stabilize $s$. Rewriting ${\cal L}^H_{\rm bos}$ in the
heterotic string frame,
\beq
\label{dilis}
e^{-2\phi}=s ~~~~~ ~~~ {\rm and }~~~~ ~~~ g^{\rm str}_{\mu\nu}=
{2\kappa^2\over\alpha^\prime_H}e^{-2\phi} g_{\mu\nu},
\eeq
we obtain
\beq
\label{stframe}
{\cal L}^H_{\rm str} = {e^{-2\phi}\over\alpha^\prime_H}
\left [-eR+4e(\partial_\mu\phi)(\partial^\mu\phi) + {2e\over
\alpha^\prime_H} \right]~;
\eeq
it is easy to show that the  $\phi$-equation of motion is that
of a 2D-sigma-model $\beta$-function equation with $\beta_{\phi}=0$
\cite{beta} and with central charge deficit
\beq
\label{deltac}
\delta\hat c = {2\over3}\delta c = -4.
\eeq
In the string frame, a background solution (\ref{stframe}) has a flat
(sigma-model) metric ${\tilde g}^{\rm str}_{\mu\nu}=\eta_{\mu\nu}$
and a linear dilaton background on a spatial coordinate:
\beq
\label{lindil}
\tilde\phi = Q_{\mu} x^{\mu}, \qquad{\rm with} \qquad
Q^2 = {\delta\hat c\over 8\alpha^\prime_H}={1\over2\alpha^\prime_H}.
\eeq
In this background there is a shift for all boson masses,
$M_B^2\rightarrow M_B^2+Q^2=M_B^2+m_{3/2}^2$ because of the
non-trivial  dilaton. The fermionic masses are also shifted because
of the temperature.
As a result, the perturbative heterotic mass spectrum
shows, fermion--boson mass degenerany due to a residual supersymmetry
existing in this background \cite{AK,ADK}. At the non-perturbative
level, however, this degeneracy is lost in the non-perturbative
massive
sector of the theory, although the ground state remains
supersymmetric
\cite{ADK}. Thus, the high-$T$ phase is expected to be stable in the
weak
coupling heterotic regime, because of the $N=2$ residual
supersymmetry.
The reader can find more details at
this point in Ref. \cite{ADK}.

\section{The high-$T$ non-critical string}
\vskip .1cm

As we discussed above, the high-$T$ phase of $N=4$ strings is
described by a non-critical string with central charge deficit
$\delta {\hat c}=-4$, provided the (six-dimensional) heterotic string
is in the weakly coupled regime, $\lambda_H\le ({\sqrt 2}+1)/2$.
One possible description is in terms of the (5+1) super-Liouville
theory compactified (at least) on $S^1_R$, with radius
fixed at the fermionic point $R=\sqrt{\alpha^\prime_H/2}$. The
perturbative stability of this ground state is guaranteed when there
is at least $\rm {\cal N}_{sc}=2$ superconformal symmetry on the
world-sheet, implying at least $N=1$ supersymmetry in space-time.

An explicit example with $\rm {\cal N}_{sc}=4$ superconformal was
given in Refs. \cite{K, AFK, ADK}. It is obtained when, together with
the temperature $S^1_R$, there is an additional compactified
coordinate
on $S^1_{R_6}$, with radius fixed at the fermionic point
$R_6=\sqrt{\alpha^\prime_H/2}$. These two
circles are equivalent to a compactification on $[SU(2)\times
SU(2)]_k$ at the limiting value of level $k=0$. Indeed, at $k=0$,
only the six world-sheet fermionic coordinates survive
describing a ${\hat c}=2$ system (with an $SO(4)_{k=1}$ current
algebra)
instead of ${\hat c}=6$ of $k\to\infty$, consistently with the
decoupling of four supercoordinates, $\delta {\hat c}=-4$.
The central-charge deficit is
compensated by the linear motion of the dilaton associated to the
Liouville field, $\phi=Q^\mu x_\mu$ with $Q^2=1/(2\alpha^\prime_H)$,
so that $\delta {\hat c}_L=8Q^2\alpha^\prime_H=4$.

Using the techniques developed in Ref. \cite{ABK, AFK}, we derive in
Refs. \cite{ADK} the one-loop (perturbative) partition function in
the
high-$T$ heterotic phase. Here I stress some of our results. More
details will appear in Ref. \cite{ADK}.

\begin{itemize}
\item
The initial $N=4$ supersymmetry is reduced in the high-$T$ heterotic
phase to $N=2$. This agrees with the effective field theory analysis
of the
high-$T$ phase. The (perturbative) bosonic and fermionic mass
fluctuations are degenerate because of the remaining $N=2$
supersymmetry (${\cal N}_{\rm sc}=4$ superconformal on the
world-sheet).
\item
The spectrum of the theory contains two sectors, $h=0$ and $h=1$:
the $h=0$ sector has no  massless fluctuations;
the bosonic and fermionic masses (squared) are shifted by $m^2_{3/2}$
because of the dilaton background and the temperature coupling;
all masses are larger than or equal to $m_{3/2}$. This is again in
agreement with the effective theory analysis.
\item
In the $h=1$ sector there are massless excitations, as
expected from the (5+1) super-Liouville theory \cite{BG, AFK}.
\item
The $5+1$ Liouville background can be regarded as an Euclidean
5-brane solution wrapped on $S^1\times S^1$ preserving one-half
($N=2$) of the initial $N=4$ space-time supersymmetries.
\item
The massless space-time fermions coming from the $h=1$ sector are
six-dimensional space-time spinors; they are also spinors under the
$SO(4)_{k=2}$ right-moving group defined at the fermionic point of
the
$S^1_R\times S^1_{R_6}$ compactification; they are also in the vector
representation of an $SO(28)_{k=1}$ heterotic right-moving
group.
\item
The massless space-time bosons are in the same right-moving
representation, {\it e.g.} spinors under $SO(4)_{k=1}$  and
vectors under $SO(28)_{k=1}$ right-moving groups. In addition,
they are spinors under $SO(4)_{k=1}$ left-moving group.
Together with the massless fermions,
they form 28 $N=2$ hypermultiplets sitting on the special
quaternionic space:
$$
{\rm \cal H}={SO(4,28)\over SO(4)\times SO(28)}.
$$
\end{itemize}

The 28 massless hypermultiplets in the $h=1$ sector are the only
states that survive in the zero-slope limit. Their effective field
theory is described by an $N=2$ 4D-sigma-model on a hyper-K\"ahler
manifold $\rm \cal{K}$, which is obtained from ${\rm \cal H}$ in the
decoupling limit of gravity. This topological theory corresponds to
the infinite temperature limit of the $N=4$ strings after the
heterotic Hagedorn phase transition \cite{ADK}.

Although the $5+1$ Liouville background is perturbatively stable,
owing
to the ${\rm \cal N}_{sc}=4$ superconformal symmetry, its stability
is not ensured at the non-perturbative level when the heterotic
coupling is large:
\beq
\label{hetcoup}
g^2_H(x_{\mu})=e^{2(\phi_0 - Q^{\mu}x_{\mu}) } ~>~
g^2_{\rm crt}={\sqrt{2}+1 \over 2\sqrt{2}} \sim 0.8536 ~.
\eeq
As we explained above, the high-$T$ heterotic phase exists only if
$g^2_H(x_{\mu})$ is lower than a critical value that separates the
heterotic and Type II high-$T$ phases. Thus one expects a domain-wall
in space-time, at $Q^{\mu}x_{\mu}^{0}=0$, separating these two
phases: $g^2_H(Q^{\mu} x^0_{\mu})\sim g^2_{\rm crt}$. The domain wall
problem can be avoided by replacing the $(5+1)$ super-Liouville
background by a more appropriate one having the same superconformal
properties, ${\rm \cal N}_{\rm sc}=4$, obeying however the additional
perturbative constraint $g^2_H(x_{\mu}) \ll g^2_{\rm crt}$ in the
entire space-time.

Exact superstring solutions based on gauged WZW two-dimensional
models with $\rm {\cal N}_{sc}=4$ superconformal symmetries have been
studied in the literature \cite{KPR, K, AFK, KKL}. In Ref.
\cite{ADK},
various candidate backgrounds with $\delta \hat{c}=-4$ are
considered.
The first one is that of $(5+1)$ super-Liouville ($\delta{\hat
c}_{L}=4$)
already examined above. Another class of candidate backgrounds
is based on the non-compact parafermionic spaces $\rm {\cal M}^4_P$
that are described by gauged WZW models:
$$
{\rm {\cal M}^4_p}~=~\left[{SL(2,R) \over U(1)_{V,A} }\right]_{k=4}
\times
\left[{SL(2,R) \over U(1)_{V,A} }\right]_{k=4} \times
 U(1)_{R^2={\alpha^\prime_H}/2} \times
U(1)_{R_6^2={\alpha^\prime_H}/2}
$$
\beq
\equiv \left[{SL(2,R) \over U(1)_{V,A} }\right]_{k=4} \times
\left[{SL(2,R) \over U(1)_{V,A} }\right]_{k=4}
\times SO(4)_{k=1}\,;
\eeq
the indices $A$ and $V$ stand for the ``axial" and ``vector" WZW
$U(1)$ gaugings. Then, many backgrounds can be obtained by marginal
deformations of the above, which preserve at least $\rm{\cal
N}_{sc}=2$, or also by performing $\cal S$- or $\cal T$-duality
transformations on them.

As already explained, the appropriate background must verify the
weak-coupling constraint (\ref{hetcoup}). This weak-coupling
limitation is achieved in the
``axial" parafermionic space ${\rm {\cal M}^4_p}(\rm axial)$.
Indeed, in this background,
$g^2_H(x_{\mu})$ is bounded in the entire non-compact
four-dimensional space, with coordinates $x_{\mu}=\{z,z^*,w,w^*\}$,
provided the initial value of $g^2_0=g^2_H(x_{\mu}=0)$ is small:
\beq
{1\over g^2_H(x{\mu})}=e^{-2\phi}={1\over g^2_0}~(1+zz*)~(1+ww*)~\ge
{}~{1\over g^2_0} ~\gg ~{1\over g^2_{\rm crt}}.
\eeq
The metric of this background is everywhere regular,
\beq
ds^2={4dzdz^*\over 1+zz^*} +{4dwdw^*\over 1+ww^*}\,\,,
\eeq
while the Ricci tensor and the scalar curvature
$$
R_{z\,z^*}={1\over (1+zz^*)^2}~, {\rm~~~~~~} R_{w\,w^*}={1\over
(1+ww^*)^2}\,, ~~~~~~ R={1\over 4(1+zz^*)}+{1\over 4(1+ww^*)}
$$
vanish for asymptotically large values of $|z|$ and $|w|$
(asymptotically flat space). This space has maximal curvature when
$|z|=|w|=0$. This solution has a behaviour similar to the one of the
Liouville solution in the asymptotic regime $|z|,~|w|~\rightarrow
{}~\infty$. In this limit, the dilaton $\phi$ becomes linear when
expressed in terms of the flat coordinates $x_i$:
$$
\phi=-\Re[{\rm log}z]-\Re[{\rm log}w] = -Q^1|x_1|-Q^2|x_2|,
$$
\beq
x_1=-{\Re}[{\rm log}z], \quad
x_2=-{\rm Re}[{\rm log}w], \quad
x_3={\rm Im}[{\rm log}z], \quad
x_4={\rm Im}[{\rm log}w].
\eeq
In the large- $|z|$ and $|w|$ limit, $\rm {\cal M}^4_P$ is flat with
$ds^2=4(dx_i)^2$. The important point here is that for large values
of $|x_1|$ and $|x_2|$, $\phi \ll 0$, in contrast to the Liouville
background in which $\phi=Q^1x_1+Q^2x_2$ and therefore  the dilaton
is becoming positive and arbitrarily large in one half of the space,
thus violating the weak-coupling constraint (\ref{hetcoup}).

We then conclude that the high-$T$ heterotic phase is well described
by
the ${\rm {\cal M}^4_p}(\rm axial)$. This space is $N=2$
supersymmetric and stable when $g^2_0 \ll 1$, since it is everywhere
perturbative with degenerate massive bosonic and fermionic
fluctuations.
The non-perturbative states are superheavy and decouple in the
limit of vanishing coupling $g^2_0\rightarrow \infty$.

On the heterotic or type IIA side, the high-temperature limit after
the high-$T$ heterotic phase transition  corresponds to a topological
$N=2$ supersymmetric theory described by a 4D-sigma-model on a
non-trivial hyper-K\"ahler manifold. On the type IIB side, on the
other hand, the high-$T$ phase corresponds to a tensionless
string defined by a limit that generalizes the large-$N$ limit of
Yang--Mills theory, $\alpha^\prime_H \rightarrow \infty$,
$\lambda_H\rightarrow 0$,
with $\alpha^\prime_H\lambda_H$ fixed \cite{ADK}. It is very
interesting to study further
the properties of these theories that describe the high-$T$ phase of
string theory, in more general
compactifications with lower number of supersymmetries, and to study
their possible cosmological implications.

\vskip .4cm
\centerline{\bf Acknowledgements}
\vskip .1cm

These notes are based on work done in collaboration with I.
Antoniadis and J.-P. Derendinger. I would like to thank the
organizers of the ``6th Hellenic School and Workshops on Elementary
Particle Physics'' and the EEC contract TMR-ERBFMRX-CT96-0045 and
TMR-ERBFMRX-CT96-0090 for partial financial support.

\end{document}